\documentclass{article} 
\usepackage{microtype}
\usepackage{graphicx}
\usepackage{subfigure}
\usepackage{booktabs} 
\usepackage[table, usenames,dvipsnames]{xcolor}
\usepackage{soul} 
\usepackage{bm}
\usepackage{comment}
\usepackage{csquotes}
\MakeOuterQuote{"} 

\usepackage{hyperref}

\usepackage[accepted]{icml2024}

\usepackage{amsmath}
\usepackage{amssymb}
\usepackage{mathtools}
\usepackage{amsthm}
\usepackage{physics}
\usepackage{enumitem}

\usepackage[capitalize,noabbrev]{cleveref}
\crefname{theorem}{Thm.}{Thms.}
\crefname{corollary}{Cor.}{Cors.}
\crefname{bclaim}{Claim}{Claims}
\crefname{proposition}{Prop.}{Props.}
\crefname{section}{Sec.}{Secs.}
\crefname{appendix}{App.}{Apps.}
\crefname{axiom}{Axiom}{Axioms}

\theoremstyle{plain}
\newtheorem{theorem}{Theorem}[section]

\theoremstyle{definition}

\newtheorem{assumption}[theorem]{Assumption}
\theoremstyle{remark}

\usepackage[textsize=tiny]{todonotes}

\icmltitlerunning{Geometric Dynamics of Signal Propagation Predict Trainability of Transformers}

\definecolor{brightmaroon}{rgb}{0.76, 0.13, 0.28}
\newif\ifcommentflag
\commentflagfalse

\newcommand{\AC}[1]{\ifcommentflag {\color{brightmaroon}{#1}}\fi}

\newcommand{\TT}[1]{\ifcommentflag {\color{BrickRed}{#1}}\fi}

\definecolor{brickred}{rgb}{0.8, 0.25, 0.33}
\usepackage{xspace}

\newcommand{\MLP}{{\operatorname{MLP}}}
\newcommand{\Att}{{\operatorname{ATT}}}
\newcommand{\Norm}{{\operatorname{Norm}}}
\newcommand{\alphat}{\tilde{\alpha}}

\usepackage{bbold}
\newcommand{\Id}{\mathbb{1}} 
\newcommand{\EE}{\mathbb{E}}
\def\T{\intercal} 

\begin{document}

\twocolumn[
\icmltitle{Geometric Dynamics of Signal Propagation Predict Trainability of Transformers}



\icmlsetsymbol{equal}{*}

\begin{icmlauthorlist}
\icmlauthor{Aditya Cowsik}{phys}
\icmlauthor{Tamra Nebabu}{phys}
\icmlauthor{Xiao-Liang Qi}{phys}
\icmlauthor{Surya Ganguli}{applied}

\end{icmlauthorlist}

\icmlaffiliation{phys}{Department of Physics, Stanford University, Stanford, CA 94305}
\icmlaffiliation{applied}{Department of Applied Physics, Stanford University, Stanford, CA 94305}

\icmlcorrespondingauthor{Aditya Cowsik}{acowsik@stanford.edu}

\icmlkeywords{Machine Learning, ICML}

\vskip 0.3in
]



\printAffiliationsAndNotice{}  

\begin{abstract}
We investigate forward signal propagation and gradient back propagation in deep, randomly initialized transformers, yielding simple necessary and sufficient conditions on initialization hyperparameters that ensure trainability of deep transformers. 
Our approach treats the evolution of the representations of $n$ tokens as they propagate through the transformer layers in terms of a discrete time dynamical system of $n$ interacting particles. 
We derive simple update equations for the evolving geometry of this particle system, starting from a permutation symmetric simplex. 
Our update equations show that without MLP layers, this system will collapse to a line, consistent with prior work on rank collapse in transformers. 
However, unlike prior work, our evolution equations can quantitatively track particle geometry \TT{This is not a real geometry but an emergent "effective" geometry in the ensemble average.} in the additional presence of nonlinear MLP layers, and it reveals an order-chaos phase transition as a function of initialization hyperparameters, like the strength of attentional and MLP residual connections and weight variances. 
In the ordered phase the particles are attractive and collapse to a line, while in the chaotic phase the particles are repulsive and converge to a regular $n$-simplex. 
We analytically derive two Lyapunov exponents: an angle exponent that governs departures from the edge of chaos in this particle system, and a gradient exponent that governs the rate of exponential growth or decay of backpropagated gradients. We show through experiments that, remarkably, the final test loss at the {\it end} of training is well predicted just by these two exponents at the {\it beginning} of training, and that the simultaneous vanishing of these two exponents yields a simple necessary and sufficient condition to achieve minimal test loss.

\end{abstract}

\section{Introduction and Related Work}

Deep transformers \citep{vaswani2017attention} with many layers have been incredibly successful in a variety of domains from NLP \cite{devlin2018bert,brown2020language}, to vision \cite{dosovitskiy2020image,khan2022transformers,han2022survey,arnab2021vivit}. Such transformer layers involve many components, including attention, a nonlinear MLP layer, and residual connections, our understanding of how signals propagate through many such complex layers, even at initialization, is still rudimentary. 
It is unclear how to quantitatively describe this signal propagation and its dependence on hyperparameters, as well as how to use any such quantitative description to rationally choose good initialization hyperparameters that ensure good final test loss.  Here we derive an analytic description of both forward signal propagation of tokens through transfomers, as well as the back-propagation of gradients, and we experimentally show that two simple properties of this signal propagation at {\it initialization} are sufficient to predict test loss at the {\it end} of training. 

Our work extends to transformers a body of work that developed quantitative theories of forward and backward signal propagation through pure deep MLP networks \cite{Saxe2014-ch,poole2016exponential,Schoenholz2017-fi,Pennington2017-za,Pennington2018-fy, doshi2023critical, he2022autoinit}. In particular \cite{poole2016exponential} described quantitatively how the geometry of pairs of inputs changed as they propagate through the layers of a randomly initialized nonlinear MLP. This analysis revealed the existence of two distinct dynamical phases of signal propagation depending on initialization hyperparameters: ordered and chaotic. In the ordered (chaotic) phase nearby inputs converge (diverge) and backpropagated gradients vanish (explode). Initialization along a co-dimension $1$ phase boundary in hyperparameter space, i.e. at the edge of chaos, was shown to constitute a {\it single} necessary and sufficient condition on initialization hyperparameters to enable the trainability of deep MLPs \cite{Schoenholz2017-fi}. \citet{doshi2023critical} extended this to a generic case with skip connections and layer norm.

Here we show how to extend this work to transformers, which is more complex because we must track the geometry of $n$ different inputs, corresponding to $n$ tokens, as they simultaneously propagate through transformer blocks.  And interestingly, we will find not $2$ but $4$ distinct phases marked by distinct properties of forward versus backward signal propagation.  In the forward direction, we will find an order-chaos phase transition between two phases: an ordered phase where the $n$ token representations converge and collapse to a line, and a chaotic phase where the $n$ token representations chaotically repulse each other and converge to a regular $n$-simplex.  In the backward direction, we will find a different dynamical phase transition between two phases: one phase corresponding to exponentially exploding gradients across layers, and another phase corresponding to exponentially vanishing gradients. Each of these phase transitions possesses a co-dimension $1$ phase boundary in initialization hyperparameter space, but these two phase boundaries are {\it not identical}, as in the case of pure MLPs.  Thus depending on initialization hyperparameters, forward signal propagation can either be ordered or chaotic, and backward gradient propagation can either be vanishing or exploding, yielding $4$ possible dynamical phases of signal propagation in transformers.  In contrast, because forward and backward phase boundaries are identical in MLPs, they only exhibit two distinct phases.  We will further find that initializing hyperparameters at the intersection of these two phase boundaries constitutes a simple necessary and sufficient condition for ensuring low final test loss at the end of training.  Moreover we will derive two Lyapunov exponents that measure departures from each phase boundary in initialization hyperparameter space, and show that a combination of just these $2$ numbers can, surprisingly, quantitatively predict the test loss at the {\it end} of training.

In related work on transformers, \citet{dinan2023effective} developed a theory of hyperparameter initialization for transformers in the large width limit, though their analysis did not explicitly take the depth of the transformer into account and therefore did not analyze deep signal propagation.  Other work has analyzed rank collapse of attention matrices in deep transformers (\citet{dong2021attention, noci2022signal,noci2023shaped}), which leads to vanishing gradients. Thus, to ensure the trainability of deep transformers one must tune initialization hyperparameters to prevent rank collapse. A similar question was studied in the context of ResNets by \cite{martens2021rapid}, who proposed a means to shape the network's kernel at initialization to facilitate training. It is desirable to develop a analogous theory for deep transformers that preconditions a deep transformer for trainability. \cite{he2023simplifying} approached the question of trainability by modifying the transformer blocks. And recent work \cite{geshkovski2023emergence,Geshkovski2023-sf} provided an elegant analysis of signal propagation in deep transformers consisting of pure attention layers without an MLP.  They viewed the dynamics of $n$ token representations propagating through a sequence of attention blocks as a dynamical system of $n$ particles evolving in a $d$ dimensional embedding space. We adopt this elegant perspective, but we note that it is unclear how to easily extend their analysis methods to go beyond pure attention and include nonlinear MLP layers. In contrast, our analysis method {\it quantitatively} accounts for the composition of attention, nonlinear MLPs, and residual connections, revealing distinct phases of signal propagation that would be absent under pure attention.





\AC{
\begin{enumerate}
    \item \cite{he2023simplifying} talks about how you can simplify transformer blocks, by removing connections, performing attention/mlp in parallel etc.
    \item \cite{martens2021rapid} essentially derive the optimal shape of the activation function in a resnet by imposing certain conditions on the token angle/token norm layer-to-layer map.
    \item \cite{dinan2023effective} This is Emily, Sho, and Susan's paper on the scaling of transformers at initialization. I think they more or less consider large-width and consider an NTK perspective on it. 
    \item \cite{geshkovski2023emergence} -- This is the paper Surya sent about the infinite depth limit in pure attention transformers (with fixed attention weights) where the points limit to a polytope.
\end{enumerate}
}

In \cref{sec: setup} we introduce the transformer architecture and the random initialization ensemble that we consider.  In \cref{sec:transf_dyn} we study the typical properties of both forward and backward signal propagation in this ensemble of randomly initialized transformers, deriving analytically the locations of the order-chaos phase boundary in forward propagation, and the exploding-vanishing phase boundary in backward gradient propagation. In \cref{sec: numerical_dyn} we provide experimental tests of our theory of signal propagation at initialization using numerical experiments on actual transformers. Lastly, in \cref{sec: trainability} we demonstrate the relevance of our theory to training, by providing necessary and sufficient conditions for good trainability, and showing how to predict the final test loss using {\it only} two real numbers (corresponding to departures from the order-chaos and vanishing-exploding phase boundaries) computed at initialization. 

\section{Setup}\label{sec: setup}
\begin{figure}
    \centering
    \includegraphics[width=\columnwidth]{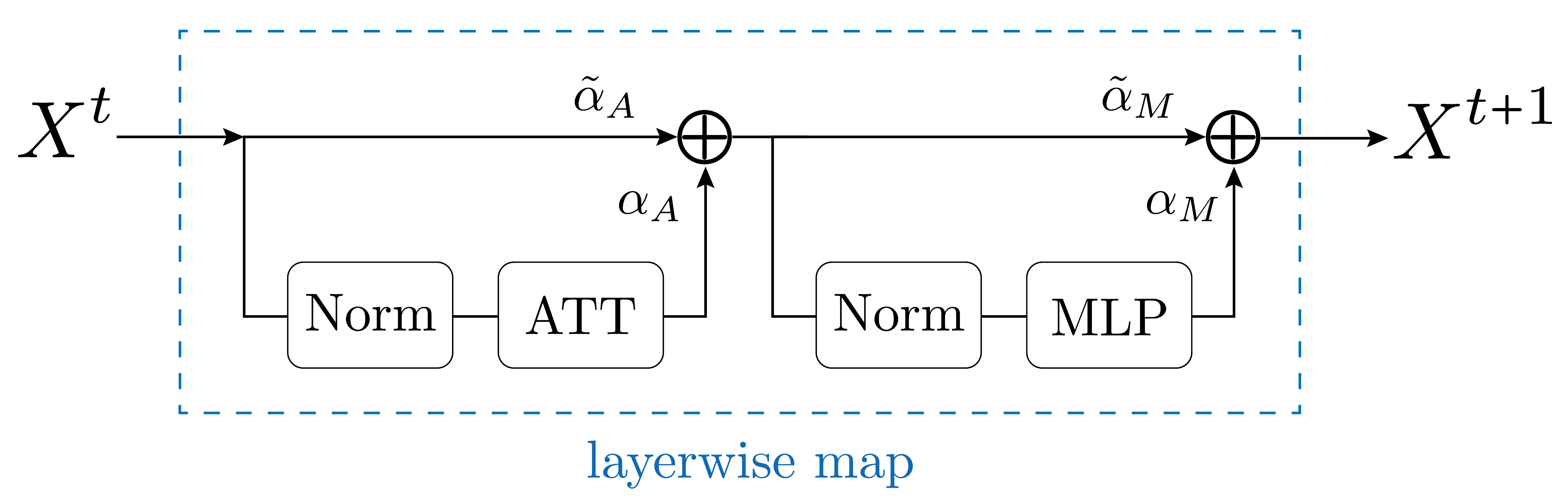}
    \caption{Schematic for the layerwise map of the transformer where $t$ is a layer index. The norm and MLP blocks operate tokenwise (i.e. act on the vector components) while the attention block operates on all of the tokens.}
    \label{fig:layerwise_map}
\end{figure}

We study a random ensemble of deep transformers acting on $n$ tokens, $X_i \in \mathbb{R}^d, ~ 1 \leq i \leq n$ where $d$ is the embedding dimension. The layerwise map we consider is composed of a single-head self-attention block followed by a tokenwise $\ell$-layer multilayer perceptron (MLP) block with a residual branch (see \cref{fig:layerwise_map}). We take $\ell = 2$ in all experiments. The three main components of the layerwise map are the attention block, the MLP layer, and the normalization. The attention block acts jointly on a set of tokens $X_i$ as
\begin{equation}
\begin{aligned}
    \Att(X)_i &= V \sum_{j} A_{ij} X_j,\\
    A_{ij} &= \frac{e^{(Q X_i) \cdot (K X_j)/\sqrt{d}}}{\sum_{k=1}^n e^{(Q X_i) \cdot (K X_k)/\sqrt{d}}}.
    \label{eq:Att_action}
\end{aligned}
\end{equation}
Meanwhile, the MLP block acts tokenwise as
\begin{equation}
    \MLP(X_i) = W^\ell \phi( W^{\ell-1} \cdots W^1\phi(W^0 X_i)),
    \label{eq:MLP_action}
\end{equation}
where $\phi$ denotes the elementwise non-linearity, and $W^k$ are the weight matrices. Throughout this work, we consider a bias-free MLP layer and take $\phi =\tanh$ unless otherwise stated, as a generic sufficiently non-linear activation function. This also simplifies later analysis by simplifying the behavior of the token normalization at the fixed point, though the analysis does not otherwise depend strongly on this assumption.

For the normalization block, we consider a layer-norm that acts independently on tokens and is defined by
\begin{equation}
    \Norm(X_i) = \frac{\sqrt{d}}{|X_i|_2} \cdot X_i
    \label{eq:layer_norm}
\end{equation}
so that the overall 2-norm of each token is $\sqrt{d}$. To simplify the analysis we remove the trainable parameters in this normalization. 

The overall layerwise map is defined by the composition of 
\begin{align}
    X &\mapsto \tilde{\alpha}_A X + \alpha_A \Att(\Norm(X)) \nonumber \\
    X &\mapsto \tilde{\alpha}_M X + \alpha_M \MLP(\Norm(X)).
    \label{eq:full_transformer}
\end{align}
\TT{Suggestion based on later text: The overall layerwise map $F$ is defined as follows:
\begin{align}
    F &= F_\MLP \circ F_\Att \label{eq:full_transformer}
    F_\Att &= \alphat_A \Id + \alpha \Att\\
    F_\MLP &= \alpha_M \Id + \alpha \MLP
\end{align}
}
Here $\alpha_A$ and $\tilde{\alpha}_A$ control the strengths of attention and it's residual branch respectively, and similarly $\alpha_M$ and $\tilde{\alpha}_M$ control the strength of the MLP and it's residual branch. 

We consider an ensemble of transformers with $Q,K,V,W^k$ in each layer initialized with independent Gaussian entries, so that $V_{\mu\nu} \sim \mathcal{N}(0, 1/d)$, $W^k_{\mu\nu} \sim \mathcal{N}(0, \sigma_w^2/d)$, and $\EE\left[(Q^\T K)_{\mu\nu}^2\right] = \sigma_A^2/d$. 

We will think of the $n$ token representations $X_i$ in any layer $t$ as a set of $n$ particles evolving in the $d$ dimensional embedding space. At any layer $t$, the geometry of this set of $n$ particles can be described by the $n$ by $n$ matrix of dot products  
\begin{equation}
    C_{ij} \equiv X_i \cdot X_j.
    \label{eq:token_angle_matrix}
\end{equation}
Due to the randomness in the initialization $Q,K,V,W^k$, for any initial dot product matrix at the input layer, the dot product matrix at subsequent layers will be random matrices. However, in the limit of large $d$ and $n$, we expect these random matrices to concentrate about their expectation over the random parameters $Q,K,V,W^k$.  Therefore we can study the typical evolution of the geometry of these $n$ particles by deriving deterministic update equations for how the {\it expected value} of the $n$ by $n$ dot product matrix $C$ evolves across layers. This is one of our fundamental goals in \cref{sec:transf_dyn}.   

\section{Theory of transformer signal propagation}\label{sec:transf_dyn}

In \cref{subsec:attevol} through \cref{subsec:fixedpoints} we study the forward propagation of signals through a transformer, revealing an order-chaos phase transition where the $n$ token representations, as they evolve through the layers, can be thought of as attractive particles that collapse to a line in the ordered phase, or repulsive particles that converge to an $n$-simplex in the chaotic phase. In \cref{subsec:gradprop} we study backpropagation of gradients, and derive a different phase transition between vanishing and exploding gradients.

\subsection{Evolution Under Attention}
\label{subsec:attevol}
Deriving an update equation for the expected value of $C$ across an attention layer is in general difficult because one must solve \TT{the coupled dynamics of the} for the $n(n+1)/2$ distinct elements of $C$. However, the update equation for $C$ simplifies if we make a permutation invariant assumption:  
\begin{assumption}
    \label{ass:permutation_symmetry}
    We assume a permutation invariant initial condition for the $n$ particles in the input layer such that the initial dot product matrix $C$ is given by:
    \begin{equation}
        C = \begin{pmatrix}
            q & p & p & \dots & p \\
            p & q & p & \dots & p \\
            \vdots &&&& \vdots    \\
            p & p & p & \dots & q
        \end{pmatrix}.
        \label{eq:symmetric_ansatz}
    \end{equation}   
    This corresponds to a configuration of $n$ particles lying at the vertices of a regular $n$-simplex with a cosine angle of $p/q$ for all particle pairs, and squared norm $q$ for all particles. 
\end{assumption}
Now because of the permutation symmetry of acausal attention, if the token or particle configuration obeys the permutation symmetric assumption for $C$ in  \cref{eq:symmetric_ansatz} in any layer, then the expected value of $C$ in the next layer, denoted by $C'$, will also obey permutation symmetry, and will be characterized by new values of the diagonal element $q'$ and off diagonal element $p'$.  Thus under permutation symmetry, we have reduced the problem of tracking $n(n+1)/2$ dot products to tracking only $2$ numbers, and our goal is now to compute this update map $(q,p) \rightarrow (q',p')$. 


Additionally, below we will provide evidence that this permutation symmetric configuration is locally attractive, by (1) giving an example in which a particle configuration with broken permutation symmetry converges under iterations of the update map to the permutation symmetric configuration; and (2) showing that numerical simulations of updates of the entire matrix $C$ do not depart substantially from permutation symmetry, when starting from a symmetric configuration.  However, the analysis of the temporal evolution of permutation symmetric configurations alone will suffice to reveal and characterize the properties of an order-chaos transition.


We first determine how $q$ and $p$ update under the attention map. Let $Y_i \equiv \Norm(X_i)$ denote the normalized tokens.  After attention, the updated matrix, $C'$, is
\begin{equation}
\label{eq: C_update}
    C'_{ij} = \sum_{k, l= 1}^n A_{ik} C_{kl} A_{jl},
\end{equation}
where we have already taken the average over $V$, which drops out as $\EE{\left[V^\T V\right]} = \Id$. 
Computing $\EE \left[C'\right]$ is difficult because it involves the second moment of $A$, which itself is a softmax. We therefore make a second assumption,


\begin{assumption}
    \label{ass:annealed}
    The denominator of $A_{ij}$ in \cref{eq:Att_action} concentrates sufficiently about its mean, and is uncorrelated from the numerator so that
    \begin{equation}
        \mathbb{E}\left[A_{ij}A_{kl}\right] \approx \frac{\mathbb{E}\left[e^{(Y_i^\T Q^\T K Y_k + Y_j^\T Q^\T K Y_l)/\sqrt{d}}\right]}{\sum_{r, s} \mathbb{E}\left[e^{(Y_i^\T Q^\T K Y_r + Y_k^\T Q^\T K Y_s)/\sqrt{d}}\right]}.
        \label{eq:att_expectation}
    \end{equation}
\end{assumption}
This assumption is valid if there are sufficiently many tokens and no one term in the softmax dominates too strongly.

When $d$ is large, the entries of the product $Q^\T K$ converge to Gaussian random variables by the central limit theorem. Distinct entries of $Q^\T K$ are uncorrelated since any distinct pair of entries corresponds to the dot product between different rows or columns of $Q$ or of $K$, so their product has zero expectation. Therefore we finish the calculation with the assumption:
\begin{assumption}
    $Q^\T K$ has i.i.d. Gaussian entries with mean zero and variance which we define to be $\sigma_A^2/d$. This is equivalent to choosing the entries of $Q,K$ to have standard deviation $\sqrt{\sigma_A/d}$ and applying the central limit theorem at large $d$.
\end{assumption}
Computing the expectation in \cref{eq:att_expectation}, and simplifying by noting that the tokens all have the same norm $\sqrt{d}$, we find that 
\begin{equation}\label{eq: A_correl}
\begin{aligned}
    \mathbb{E}\left[A_{ij}A_{kl}\right] &\approx \frac{e^{\sigma_A^2(Y_i\cdot Y_k)(Y_j \cdot Y_l)/d^2}}{\sum_{r, s} e^{\sigma_A^2 (Y_i\cdot Y_k)(Y_r \cdot Y_s)/d^2}}.
\end{aligned}
\end{equation}
\AC{Should redo this computation after removing the average token vector. This will reduce the fluctuations and the correlation between the numerator and denominator!}
Now that we have an expression for $\EE \left[{A_{ij}A_{kl}}\right]$ in terms of $C$, we can compute the ensemble averaged output token matrix $\EE [C']$ for the permutation symmetric ansatz in \cref{ass:permutation_symmetry}. Performing the sum in \cref{eq: C_update} we find that
\begin{equation}
    \EE[C'_{ij}] = \frac{d}{q} \cdot \begin{cases}
        \frac{q + p (n-1)e^{\sigma_A^2 (p/q-1)}}{1+(n-1) e^{\sigma_A^2(p/q-1)}} &i = j\\
        \frac{q + p (n-1)e^{\sigma_A^2 (p/q)(p/q-1)}}{1+(n-1) e^{\sigma_A^2(p/q)(p/q-1)}} &i\neq j
    \end{cases}
    \label{eq:c_update}
\end{equation}
Here the first case yields $q'$ while the second case yields $p'$. 
Note that when $\sigma_A^2$ is sufficiently small, both $p'$ and $q'$ reduce to $pd/q + O(n^{-1})$. \TT{I think this requires large $n$, and $d<n$.} This means that the output of attention is close to rank-1 regardless of the input token geometry. On the other hand consider the case where $p = 0$ and $\sigma_A^2$ is large. In this setting the argument of the exponential in the first case becomes large and negative, leading to $q' = q$, while it still vanishes in the second case, leading to $p' = d/n$. In this case the rank of the output is large because fluctuations in the attention matrix have grown large enough to focus on individual tokens rather than averaging them all. This recovers a result in \citet{dong2021attention} in a straightforward way, without any complex path analysis, and extends it to a large-fluctuation regime.

Next, incorporating the residual connection is straightforward 
because $\Att(X)$ and $X$ are uncorrelated due to the presence of 
the random value matrix  $V$ in $\Att(X)$. Therefore the update for the token angles with a residual connection
($\tilde{\alpha}_A \neq 0$) is simply
$\tilde{\alpha}_A^2 C + \alpha_A^2 C'$. 
This induces the following update map $(q,p) \rightarrow (q',p')$ for token geometry under one layer of attention with a residual connection:
\begin{equation}
    F_\text{Att}(q, p) \equiv \tilde{\alpha}_A^2 (q, p) + \alpha_A^2 (q', p'),
    \label{eq:attresupate}
\end{equation}
with $q'$ and $p'$ given by the diagonal and off-diagonal entries of $C'$ in \cref{eq:c_update}.

Unlike prior work, this explicit form allows us to move forward and quantitatively describe the {\it joint} effect of both attention and nonlinear MLP blocks.

\AC{need to discuss $\tilde{\alpha}^2 = 1-\alpha^2$}
\subsection{Evolution Under MLP}
Now we focus on how the MLP layers update $C_{ij}$. \citet{poole2016exponential} derived how an MLP updates $C$, though without residual connections. We briefly recall the form of their calculation.  Consider just two input vectors $X_1$ and $X_2$ to a given MLP layer, which have initial dot products $|X_1|^2 = |X_2|^2 = q$ and $X_1 \cdot X_2 = p$, which then propagate through a single random MLP layer to new activations $X'_1$ and $X'_2$. \citet{poole2016exponential} provided explicit formulas for the expected values of the dot products $|X'_1|^2 = |X'_2|^2 = q'$ and $X_1 \cdot X_2 = p'$, in terms of the MLP nonlinearity $\phi$, weight variance $\sigma_w^2/d$, bias variance $\sigma^2_b$, and the initial dot products $q$ and $p$. In the large dimension $d$ limit, the dot products concentrate about their expected values, and the formulas in \citet{poole2016exponential} are asymptotically exact.  The update equation for expected dot products revealed an order to chaos transition for most bounded nonlinearities (like $\phi=\tanh$ considered here) as $\sigma^2_w$ increases for any fixed $\sigma^2_b$.  For large (small) $\sigma^2_w$, the evolution is chaotic (ordered) with nearby inputs diverging (converging).  

It is straightforward to extend the results of \citet{poole2016exponential} to residual connections (see \cref{app:mlp_update} for details), since, as in the case of attention, $\MLP(X)$ and $X$ are uncorrelated due to the random weight matrices in $W^k$ in $\MLP(X)$.  Also the update map for $q$ and $p$ through multiple MLP layers simply follows from repeated composition of the single layer update map. In parallel to \eqref{eq:attresupate}, we denote by $F_{\text{MLP}}$ the update map $(q,p) \rightarrow (q',p')$ corresponding to an $\ell$-layer MLP with an overall residual branch. 

\AC{Need to include the definition of $f$ (from the gradient section) here as well}

\AC{Need to finish this off by discussing that we converge quickly in norm, and the angle is the slow variable, and write the update equation and discuss how $\sigma_w$ has an effect to increase the expansion}

\subsection{Fixed Points of the Update Map}
\label{subsec:fixedpoints}
A transformer layer is the composition of an MLP block with residual and an attention block with residual, so the overall evolution of $C$ is given by $F \equiv F_{\text{MLP}} \circ F_{\text{Att}}$. Because modern transformers are deep, with many layers, we will focus on a fixed point of $F$. Computing the convergence rate to this fixed point will allow us to quantify how quickly rank collapse happens as a function of hyperparameters. 

As described above, attention tends to collapse tokens towards the mean token, whereas the MLP layer may either tend to bring nearby tokens together or move them farther apart (i.e. induce chaos) depending on the activation function, number of layers inside the MLP block, and $\sigma_w$. Note that $|\tilde{\alpha}_{A}\tilde{\alpha}_M| < 1$ is necessary for the existence of a fixed point. Otherwise the norm of the tokens will tend to grow without bound as the depth of the model increases. The early layers will therefore be more impactful than later layers at initialization. For concreteness, we choose $\tilde{\alpha}^2 = 1-\alpha^2$ for both the MLP and attention branches for the remainder of this work.\footnote{The standard setup for transformers is to keep $\alphat_A=\alphat_M=1$ while varying $\alpha_A, \alpha_M$. }

\begin{figure}
    \centering
    \includegraphics[width=\columnwidth]{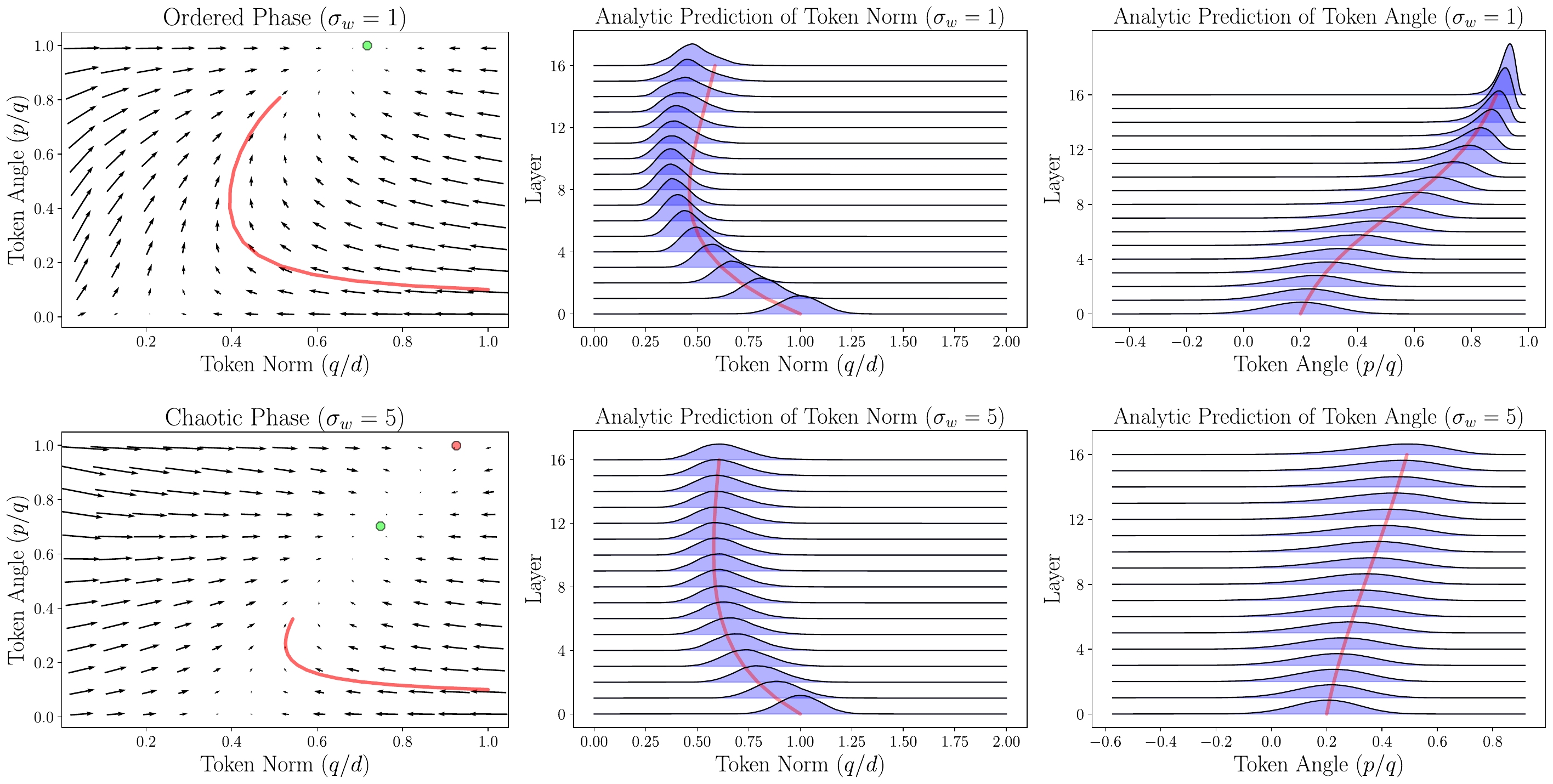}
    \caption{We show detailed agreement between our analytic theory (red curves, black arrows) and numerical simulation (blue histograms). The top row shows the token dynamics in the ordered phase where they collapse onto a line. The bottom row shows the token dynamics in the chaotic phase where they self-organize into an $n$-simplex. Our first column shows the vector field $F$ over the space of token norms $q/d$ and cosine angles $p/q$. The red curve traces 16 iterations of $F$ with $\alpha_M = \alpha_A = 8^{-1/2}$ and represents 16 layers of a transformer. We show stable (unstable) fixed points as green (red) octagons. In the second column we plot numerical distributions of token norms (blue), $q/d$, along with the analytic prediction of the expected norm (red). In the third column we similarly compare numerics to analytic predictions for the token angle, $p/q$. Our numerical simulations involve signal propagation in $16$ layer transformers with $n=256$ tokens evolving in embedding dimension $d=64$.}
    \label{fig:paper_summary}
\end{figure}

The first column of \cref{fig:paper_summary} shows iterates of $F$ starting at $q = d, p = 0$ when $\sigma_w=1$ (top row) or $\sigma_w=5$ (bottom row). The dynamics of $p$ and $q$ begin to converge to a fixed point after about 5 iterations, for a fixed $\alpha_M=\alpha_A = 1/\sqrt{8}$. For $\sigma_w$ smaller than approximately $2$, $p=q$ is a stable collapsed fixed point. This point becomes unstable for larger $\sigma_w$ with a new (stable) fixed point appearing at $p < q$, corresponding to a non-collapsed regular $n$-simplex. This relatively rapid convergence to the fixed point implies that in deep models, its character will dominate signal propagation through layers.  Furthermore, we compare our analytic predictions for the dynamics of both token norms (Fig. \ref{fig:paper_summary} second column) and token angles (Fig. \ref{fig:paper_summary} third column), to numerical simulations of token evolution through 16-layer transformers, finding an excellent match between theory and experiment.  

\begin{figure}
    \centering
    \includegraphics[width=\columnwidth]{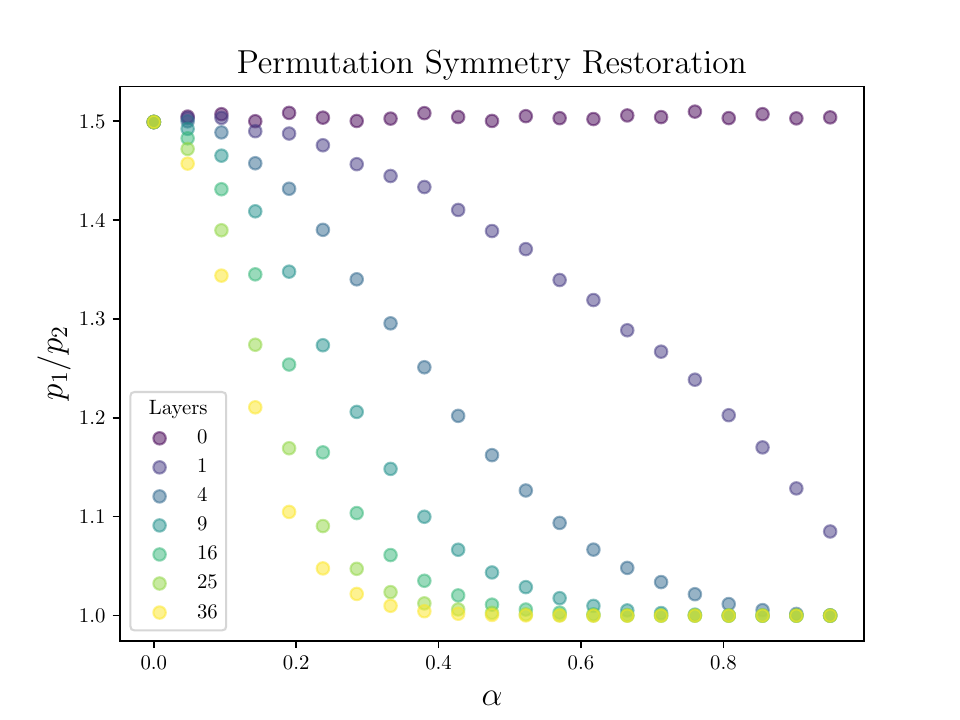}
    \caption{Statistics of tokens recover permutation symmetry (right) when it is explicitly violated in the initial conditions. We show that \cref{ass:permutation_symmetry} is robust by explicitly breaking it in the form of \cref{eq:rsb_test} so that $p_1 = 1.5 p_2 = .75d$ initially. We then compute the average $p_1 / p_2$ at several depths in a transformer, which converges to 1 for deep enough models, implying a restoration towards the form assumed.}
    \label{fig:fixed_point}
\end{figure}

Before entering into a detailed analysis of the fixed point we first verify that it is stable with respect to violations of \cref{ass:permutation_symmetry}. We choose
\begin{equation}
    C_{ij} = \begin{cases}
        q &\text{ if } i = j \\
        p_1 &\text{ if } i, j \leq \frac{n}{2} \text{ or } i, j > \frac{n}{2} \\
        p_2 &\text{ otherwise}
    \end{cases}
    \label{eq:rsb_test}
\end{equation}
for our initial condition, which explicitly violates \cref{ass:permutation_symmetry}. This condition corresponds to particles in two groups, with a smaller intra-group distance than an inter-group distance, corresponding to two offset $\frac{n}{2}$-simplices. We then pass these tokens or particles through a randomly initialized transformer with $\sigma_w = 2$. As shown in \cref{fig:fixed_point}, increasing depth results in the ratio $p_2/p_1$ tending towards 1 which indicates that permutation symmetry is restored. This robustness to permutation symmetry breaking, along with the stability of permutation symmetry indicated in our theory-experiment match in Fig.\ref{fig:paper_summary}, supports our assumption and we continue with our analysis of the permutation symmetric fixed point. 

Let $q_* = p_*$ be a fixed point of $F$ with all tokens collapsed onto one direction. Such a fixed point always exists because when all tokens are aligned, they remain aligned after one transformer block. Note that near this fixed point \cref{ass:annealed} becomes exact, because all terms in the exponential become equal. Let $dF_{q_*}$ be the 2 by 2 Jacobian corresponding to the linearization of the update map $F$ at the fixed point $q_*$.  When the largest eigenvalue of $dF_{q_*}$, which we define to be $\lambda_a+1$ (where the subscript $a$ stands for the angle between tokens), is strictly smaller than 1, convergence to the fixed point happens exponentially quickly, yielding rank collapse of attention matrices. 

Conversely, when the largest eigenvalue $1+\lambda_a$ is greater than 1, the $q_* = p_*$ fixed point becomes unstable, and another (stable) fixed point with $p<q$ emerges, corresponding to a non-collapsed fixed point in which the token geometry converges to a regular $n$-simplex. For example the first column in \cref{fig:paper_summary} demonstrates clear convergence towards such a non-collapsed fixed point when $\sigma_w=5$. This may initially seem to solve the problem of rank collapse, but it does so in a chaotic manner by exponentially amplifying small distances between pairs of nearby tokens, due to the instability of the $q^*=p^*$ fixed point. As we will show below, this chaotic amplification also impedes training. 

The boundary between collapsed and chaotic regimes occurs when the largest eigenvalue is $1$. In this case the second order term in the expansion of $F$ about the $q_* = p_*$ fixed point will control the rate of convergence to it, and so the speed of motion towards the fixed point is proportional to the \emph{square} of the remaining distance rather than the distance. Thus the distance to the fixed point decays as $O(L^{-1})$ rather than $O(e^{-\lambda_a L})$ as in the collapsed regime. This dramatic critical slowing down heralds a new large-depth limit for transformer signal propagation at initialization, in which important information about token geometry is neither erased under collapse nor scrambled due to chaos.

In summary, we can think of $\lambda_a$ as an angle Lyapunov exponent characterizing two phases: (1) an ordered or collapsing phase for $\lambda_a < 0$ in which all tokens align (with $0$ angles) exponentially fast in depth, and (2) a chaotic phase for $\lambda_a > 0$ in which nearby tokens diverge exponentially with depth and the overall token geometry converges to a regular $n$-simplex with a given nonzero angle.  Since neither property seems conducive to stable training, a natural conjecture for a good initialization is at the edge of chaos where
\begin{equation}
    \lambda_a = 0.
    \label{eq:cond_angle}
\end{equation}

\subsection{Propagation of gradients}
\label{subsec:gradprop}

A natural second condition for trainability can be derived by avoiding both exploding and vanishing gradients.
Let $X^t$ denote the tokens at transformer layer $t$, 
so that $X^0$ is the input data and $X^L$ is the output after all transformer blocks. 
Schematically, suppressing the token indices, the end-to-end input output Jacobian is 
\begin{equation}
    \frac{\partial X^L}{\partial X^0} = \frac{\partial X^L}{\partial X^{L-1}} \frac{\partial X^{L-1}}{\partial X^{L-2}} \cdots \frac{\partial X^1}{\partial X^{0}}.
    \label{eq:grad_product}
\end{equation}
Controlling gradient norms w.r.t. parameters can be done by controlling the squared Frobenius norm $\left|\frac{\partial X^L}{\partial X^0} \right|^2$ of this Jacobian.
Because this Jacobian is a product of $L$ matrices, it's norm will generically grow or decay exponentially with $L$ depending on whether the expected norm of a single block $\mathbb{E} \left|\frac{\partial X^{t+1}}{\partial X^t}\right|^2$ is bigger or less than $1$, respectively.  Using the techniques developed above, we can compute the expected norm of the end-to-end Jaobian layer by layer (see \citep{doshi2023critical} for a this analysis absent attention), making use of the identity
\begin{equation}
    \mathbb{E}\left|\frac{\partial X^L}{\partial X^0} \right|^2 = B^\T \left(\prod_{t=0}^{L-1} \mathbb{E}\left[\frac{\partial{X}^{t+1}}{\partial X^t}\otimes \frac{\partial{X}^{t+1}}{\partial X^t}\right]\right) B
    \label{eq:doubled_gradient}
\end{equation}
where $B = \sum_{i=1}^{d\cdot n} \hat{e}_i \otimes \hat{e}_i$ ($\hat{e}_i$ are the standard unit vectors). Writing the trace as an inner product of the outer product of the Jacobian with itself allows us to easily factor the expectation as a product of expectations of the layer-wise Jacobians. To simplify this calculation we focus on the $p_* = q_*$ fixed point. The single-layer expectation is
\begin{equation}
    \mathbb{E}\left[\frac{\partial{X}^{t+1}}{\partial X^t}\otimes \frac{\partial{X}^{t+1}}{\partial X^t}\right] = x \Id + y \mathcal{A} + z \mathcal{B}
    \label{eq:doubled_gradient_expectation}
\end{equation}
with
\begin{align}
    x &= \alphat_M^2 \alphat_A^2 \\
    y &= \frac{1}{q_*}\alphat_A^2 \alpha_M^2 f(\sigma_w) \\
    z &= \frac{\alpha_A^2}{q_* n^2}\left(\frac{\alpha_M^2 f(\sigma_w) d}{q^*} + \alphat_M^2 \right)
\end{align}
and the matrices $\Id, \mathcal{A}, \mathcal{B}$ obeying the algebra:
\begin{equation}
    \begin{array}{c|lcr}& \Id & \mathcal{A} & \mathcal{B}  
    \\\hline \Id & \Id & \mathcal{A} & \mathcal{B}  
    \\ \mathcal{A} & \mathcal{A} & d \mathcal{A} & d \mathcal{B}  
    \\ \mathcal{B} & \mathcal{B} & d \mathcal{B} & dn^2 \mathcal{B} \end{array}. 
\end{equation}
Finally $B^\T \Id B = d n, B^\T \mathcal{A} B = d^2 n$, and $B^\T \mathcal{B} B= d^2n^2$. These computations were carried out up to $O(d^{-1})$ corrections and further details can be found in \cref{app:gradient_calc}. Using these results it is straightforward to compute the end-to-end norm for any depth, or the scaling at infinite depth.

Typically this Frobenius norm scales exponentially, as $e^{\lambda_g L}$ where $\lambda_g$ can be thought of as a {\it gradient} Lyapunov exponent characterizing two phases: (1) a vanishing gradients phase for $\lambda_g < 0$, and (2) an exploding gradients phase for $\lambda_g > 0$. Since neither property is conducive to training, a natural conjecture for a good initialization is at the edge between these two phases in which 
\begin{equation}
    \lambda_g = 0.
    \label{eq:cond_grad}
\end{equation}

The combination of \eqref{eq:cond_angle} and \eqref{eq:cond_grad} constitute natural necessary conditions for a deep initialization to be trainable. Indeed, we will show through experiments in \cref{sec: trainability} that these conditions are necessary {\it and} sufficient. 


\section{Experimental tests of the theory}\label{sec: numerical_dyn}
Our theory provides analytic formulas for both the angle Lyapunov exponent $\lambda_a$ and gradient Lyapunov exponent $\lambda_g$, characterizing the speed of layerwise propagation of both tokens and gradients respectively, near the collapsed fixed point $q_* = p_*$ as a function of hyperparameters.  We now compare these predictions to numerical experiments.

\subsection{Theory-experiment match for the angle exponent}

\begin{figure}
    \centering
    \includegraphics[width=.95\columnwidth]{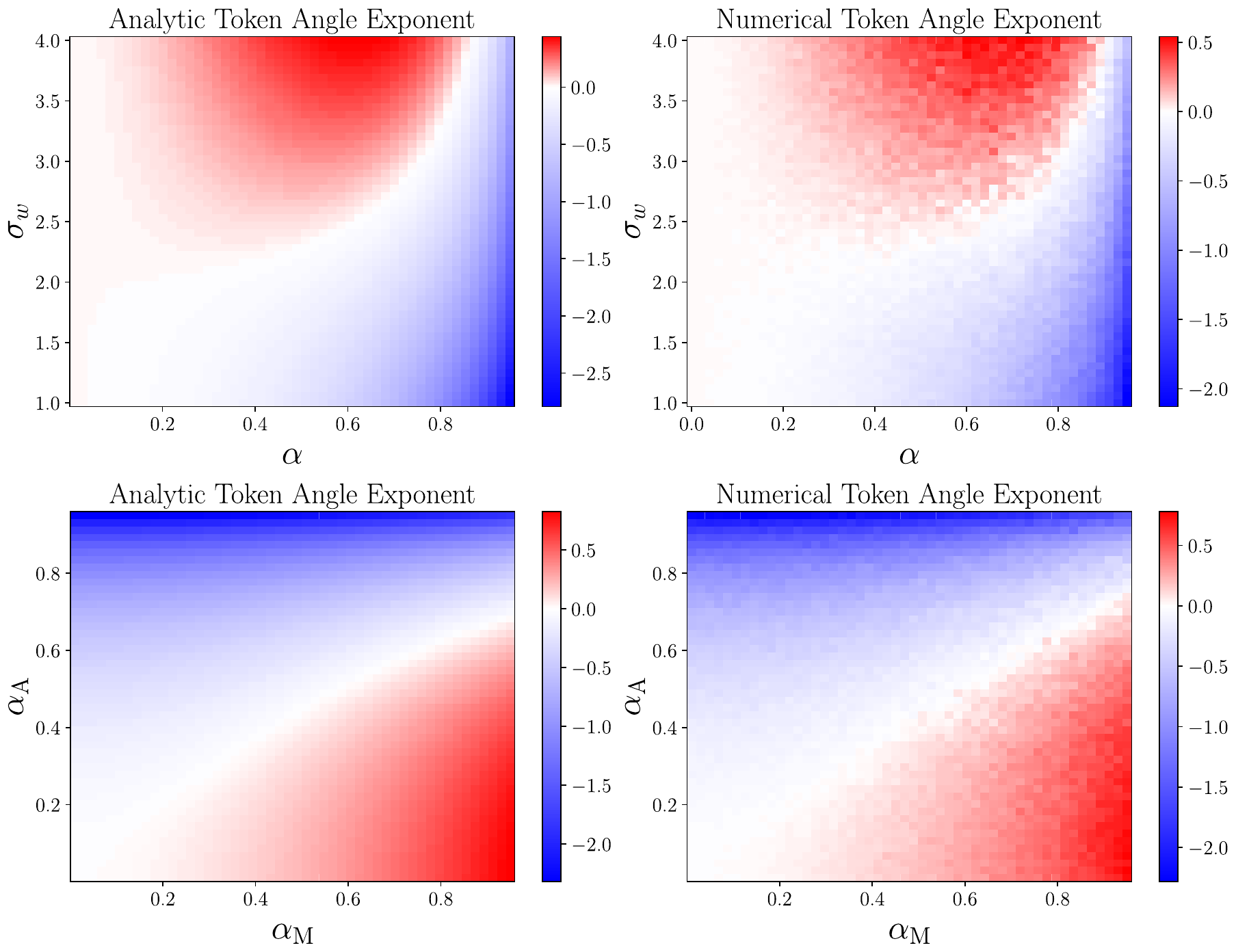}
    \caption{We show strong agreement between our analytic calculations (left column) and numerical calculations (right column) for the token angle exponent $\lambda_a$. The first row shows the phase diagram in terms of the strength of the non-residual branch $\alpha = \alpha_M = \alpha_A$ and the standard deviation of MLP weights, $\sigma_w$. The color depicts the token angle Lyapunov exponent $\lambda_a$, with a red positive (blue negative) value corresponding to the chaotic (collapsed) regime. As $\alpha$ increases, collapse due to attention strengthens, and so $\sigma_w$ must increase as well so that chaos due to the MLP can counteract attentional collapse and maintain dynamics at the edge of chaos with $\lambda_a = 0$. The second row shows $\lambda_a$ as a function of $\alpha_A$ and  $\alpha_M$ with a fixed $\sigma_w=2$. Similarly, as $\alpha_A$ increases, so must $\alpha_M$ to maintain $\lambda_a=0$, in order to balance stronger attentional collapse with stronger MLP chaotic amplification. \AC{Make the numerical figures here from a 1-layer transformer so that there is no clipping of the color axis, make sure the subfigures are all the same size.} \AC{Include $\alpha = \alpha_A = \alpha_M$ on the figure itself.}}
    \label{fig:acausal_comparison}
\end{figure}

To numerically compute the angle exponent $\lambda_a$, we randomly initialize a transformer, randomly initialize tokens obeying the permutation symmetric form of $C$ in \cref{eq:symmetric_ansatz}, pass them through one-layer of the transformer, and compute the average of the diagonal and off-diagonal entries to determine $q', p'$ respectively. From this we numerically extract the lyapunov exponent $\lambda_a$ as the logarithm of the expansion factor $\log \left[(1-p'/q')/(1-p/q)\right]$,  which characterizes the rate of exponential growth of the angle between tokens when $p$ and $q$ start near the collapsed $q^*=p^*$ fixed point. We compare this numeric value to the theoretical value of $\lambda_a$ extracted from the maximum eigenvalue of the Jacobian $dF_{q_*}$ of the update map $F$ at this fixed point. We find excellent agreement between the two values over both an $(\alpha,\sigma_w)$ and an $(\alpha_A,\alpha_M)$ slice of hyperparameter space (\cref{fig:acausal_comparison}). This agreement justifies our earlier approximations for random transformers and data.

Near $\alpha_M = \alpha_A = 0$ we find $\lambda_a \approx 0$, which provides a theoretical explanation for the success of standard initialization schemes that choose $\alpha \sim 1/\sqrt{L}$. Interestingly, away from this standard line, one can {\it also} achieve $\lambda_a=0$ by choosing hyperparameters so as to balance attention induced collapse and MLP induced chaos, as explained in \cref{fig:acausal_comparison}. 

\subsection{Theory-experiment match for gradient exponent}
\begin{figure}
    \centering
    \includegraphics[width=.95\columnwidth]{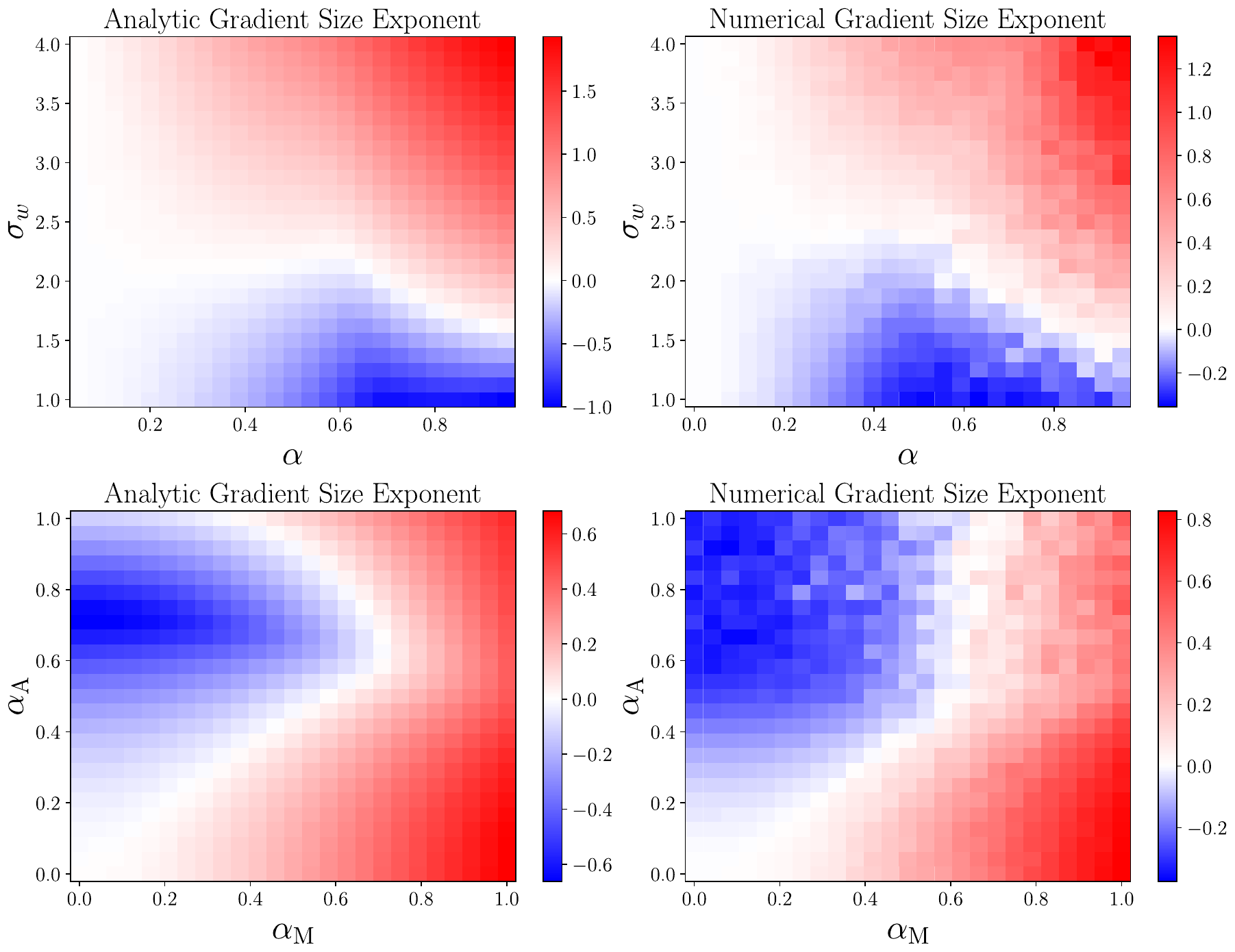}
    \caption{The gradient exponent $\lambda_g$ in a deep transformer in terms of $\sigma_w$ and $\alpha = \alpha_A = \alpha_M$ (first row) and for a fixed $\sigma_w = 2$ varying $\alpha_M, \alpha_A$ (second row). The left column shows the analytic calculation while the right shows the numerical one demonstrating a substantial agreement. Compared to the angle exponent $\lambda_a$, demanding $\lambda_g = 0$ requires a smaller $\sigma_w$, but a similar ratio of $\alpha_A$ to $\alpha_M$.}
    \label{fig:acausal_gradient}
\end{figure}

To numerically compute the gradient exponent $\lambda_g$, we measure $\left|\frac{\partial\phantom{X}}{\partial X^0} (X^L \cdot R)\right|^2$ where the components of $R$ are chosen i.i.d from a zero mean unit variance Gaussian, and $L=16$. This quantity agrees in expectation over $R$ with the norm of the end-to-end Jacobian in \cref{eq:grad_product}, and is easier to compute on hardware. 
We extract the numerical value of $\lambda_g$ from the logarithm of this quantity and compare it to the theoretical value of $\lambda_g$ obtained from the analysis in \cref{subsec:gradprop}. We find a strong match (\cref{fig:acausal_gradient}).  Our theoretical analysis clearly reproduces the kink in the behavior of the $\lambda_g=0$ contour in the $(\sigma_w, \alpha)$ plane (top row) around $(\alpha, \sigma_w) \approx (.5, 2.25)$, and the positive (negative) values of $\lambda_g$ above (below) this contour.

Also, in the $(\alpha_A,\alpha_M)$ plane with a fixed $\sigma_w = 2$ (bottom row) we find the same proportional relationship between $\alpha_A$ and $\alpha_M$ along the constraint contour $\lambda_g=0$, and also see a kink in this contour $(\alpha_M, \alpha_A) \approx (.6, .5)$ in the theoretical value (left), which differs in character from numerically estimated value (right). However we note this region of discrepancy exhibits the greatest fluctuation in the estimated value of $\lambda_g$, as reflected by the speckle in color  at large $\alpha_A$.

\subsection{Theory-experiment match for global transients.}
We re-emphasize that our theory not only predicts the behavior of the tokens near the $p=q$ fixed point, but also more globally, including the transient behavior towards the fixed point, as previously described in \cref{fig:paper_summary}, which successfully compares our theoretical predictions of the expected token norm and angle (red curves) with numerically simulated distributions of these same quantities (blue histograms).  Our theory-experiment match justifies our assumptions. In particular \cref{ass:annealed} holds throughout the phase diagram despite finite-size fluctuations in norms and angles.

We see that experimentally, that the empirical spread token norms does not appear to change as tokens pass through transformer layers, and thus our theory well models the motion of the mean with only a small bias at larger depths. The token angle distribution is more interesting, becoming something akin to a log-normal distribution (specifically $1-p/q$ looks log-normally distributed) at later depths despite it's close to Gaussian initialization. Nevertheless, our analytic prediction closely tracks the mass of the distribution, despite the change in shape of the distribution. This shows that our analysis can predict the transient dynamics of token evolution,  and is robust to changes in various details about token distributions and correlations which are not taken into account in our theory. 

\subsection{Experiments on causal attention}
\begin{figure}
    \centering
    \includegraphics[width=\columnwidth]{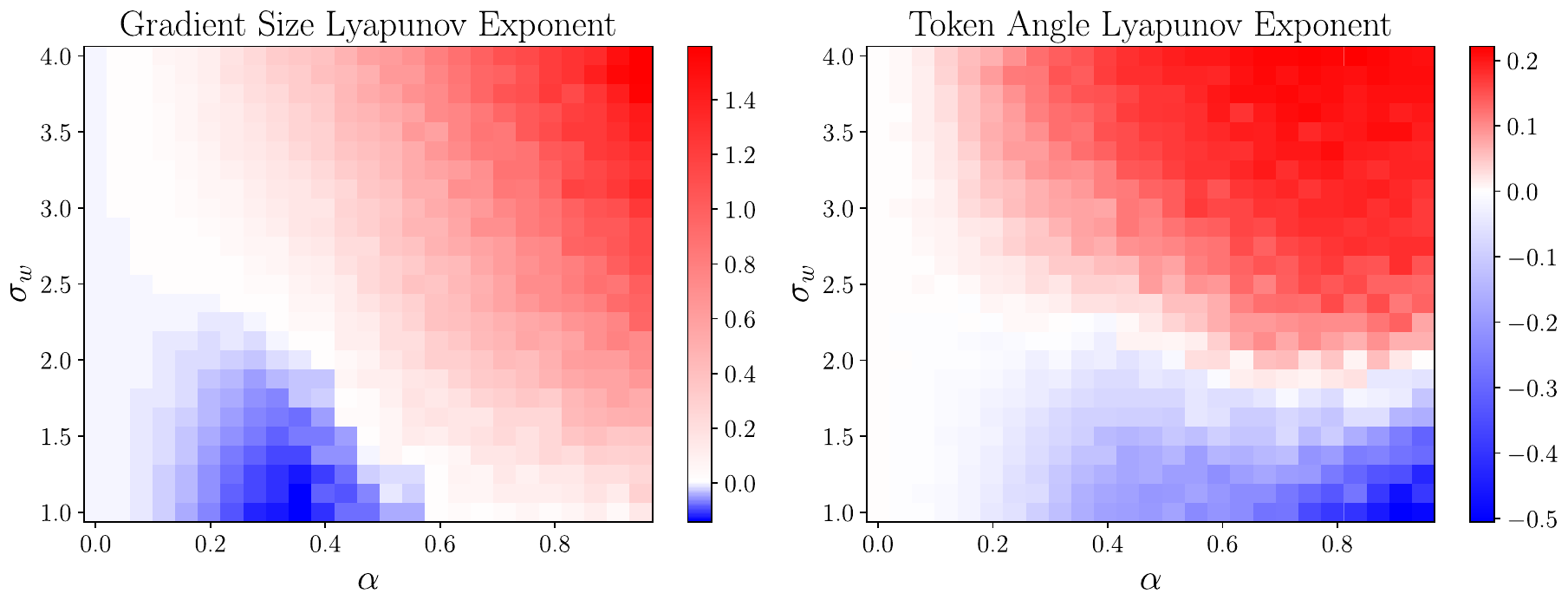}
    \caption{Exponents for causal attention as a function of $\alpha$ and $\sigma_w$.  Left: the gradient exponent $\lambda_g$ (compare to \cref{fig:acausal_gradient} top row for the acausal case). Right: the angle exponent $\lambda_a$ (compare to \cref{fig:acausal_comparison} top row for the acausal case). Upon introducing causality, both the $\lambda_g$ and $\lambda_a$ tend to increase throughout the phase diagram, pushing the constraint contours $\lambda_a=0$  and $\lambda_g=0$ towards smaller $\sigma_w$ and $\alpha = \alpha_A = \alpha_M$. The qualitative features which survive are that the $\lambda_g=0$ contour is systematically below the $\lambda_a=0$ contour, and they intersect near small $\alpha$.}
    \label{fig:with_causality_phase_diagram}
\end{figure}

While vision transformers have an all-to-all attention mechanism, transformers for language generation often have a causal mask that sets $A_{ij} = 0$ if $i < j$.  This mask ensures that tokens later in the sentence only pay attention to earlier tokens. Such masking explicitly violates permutation symmetry, and therefore makes our \cref{ass:permutation_symmetry} invalid. Therefore we computed the angle and gradient exponents $\lambda_a$ and $\lambda_g$ numerically in a transformer with causal attention to see if their dependence on hyperparameters changes substantially. Fortunately we notice that many qualitative features of the earlier analysis still survive (\cref{fig:with_causality_phase_diagram}).

The tendency for both exponents to become more positive as $\sigma_w$ increases remains as seen in \cref{fig:with_causality_phase_diagram}, which is reasonable because the MLP acts identically in both the causal and acausal attention settings. Near $\alpha = 0$ we still have both exponents near $0$ as expected, since the residual path dominates. Also the constraint $\lambda_g = 0$ lies below the constraint $\lambda_a = 0$ as $\alpha$ becomes larger. This means that as we train deeper models we must decrease $\alpha$ to satisfy both constraints closely enough for training to be feasible. 


\section{Two phase transitions predict trainability}\label{sec: trainability}

We now experimentally test whether lying on the edge of chaos phase boundary $\lambda_a=0$ {\it and} on the critical gradient propagation phase boundary $\lambda_g=0$, together constitute necessary and sufficient conditions on initialization hyperparameters to allow for successful training of deep transformer architectures. To test this hypothesis, we train a 16-layer transformer with a $\tanh$ MLP non-linearity on the Food-101 dataset \cite{bossard14}, classifying which food category corresponds to the image. We train for 15 epochs using Adam with a learning rate $\eta=3 \cdot 10^{-4}$ and then evaluate the test loss. During training we set $\alpha_A = \alpha_M = \alpha$ and sweep both $\alpha$ and $\sigma_w$, maintaining $\alphat^2 = 1-\alpha^2$. We choose a range of $\alpha \in (0, .95]$ and $\sigma_w \in [.2, 4]$ for training. For further experimental details see \cref{app:training_experiment}.  

\begin{figure}
    \centering
    \includegraphics[width=\columnwidth]{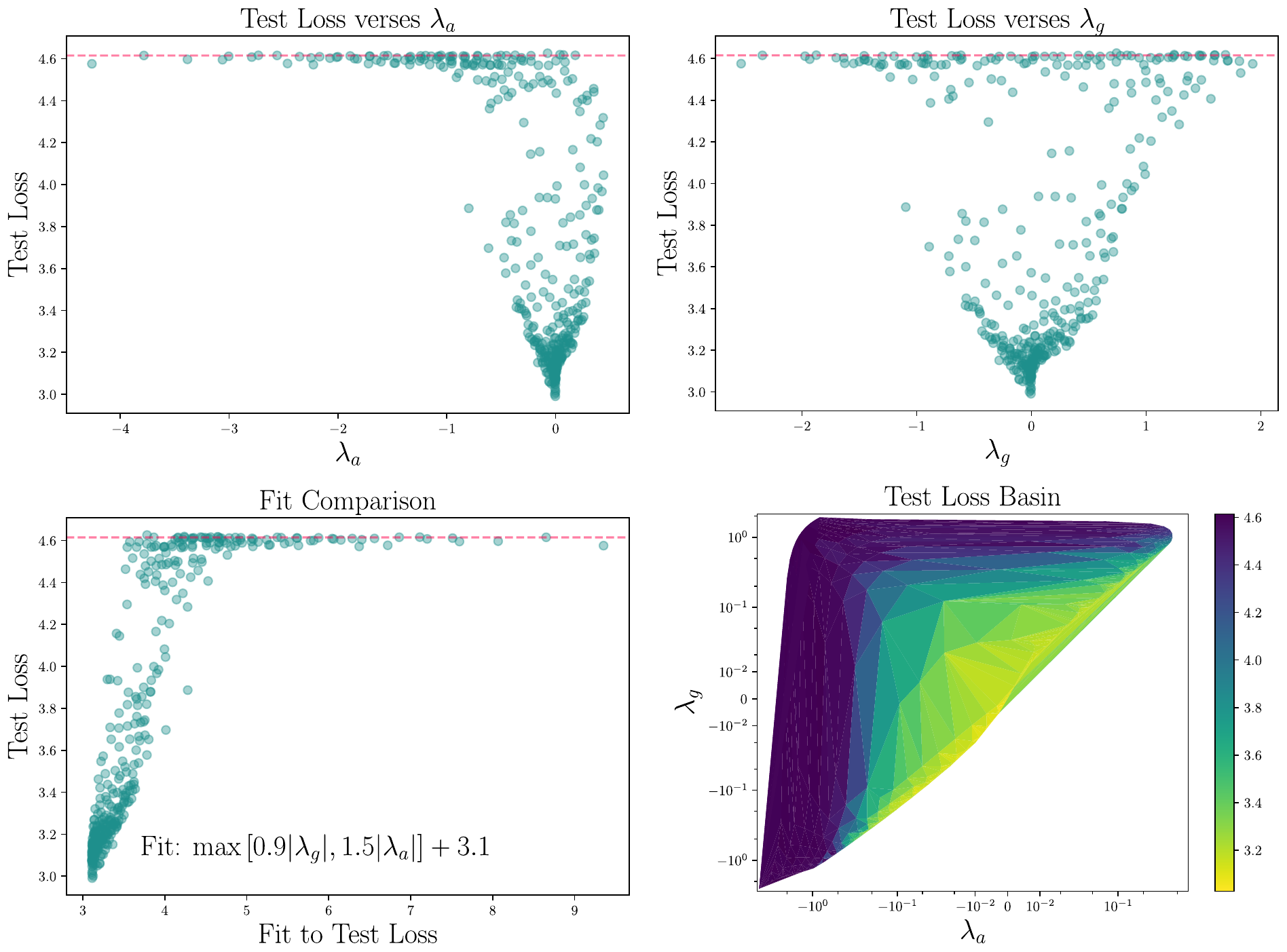}
    \vspace{-2em}
    \caption{We demonstrate empirically that the Lyapunov exponents $\lambda_a, \lambda_g$ lower bound the final test loss. The top row depicts the final test loss of our trained models as a function of $\lambda_g$ at initialization (left) and $\lambda_a$ at initialization (right). We see that the minimum test loss away from $\lambda_a = \lambda_g = 0$ increases as we move away from either of these constraints. Motivated by this we fit the final test loss taking both variables into account (bottom left) and see that we can predict the final test loss in terms of these variables. Finally we show a heatmap of the final test loss (note the non-linear scale)  which clearly demonstrates the importance of ensuring both Lyapunov exponents are near zero as deviations away from that lead to increasing test loss. Test loss has a ceiling of $\ln 101 \approx 4.6$ (dashed red line), corresponding to uniform guessing over all classes. }
    \vspace{-1em}
    \label{fig:training_comparison}
\end{figure}

This joint variation in $\alpha$ and $\sigma_w$ leads to a joint variation in the exponents $\lambda_a$ and $\lambda_g$, and in \cref{fig:training_comparison} we show the variation of test loss with $\lambda_a$ and $\lambda_g$ in various ways.  The top-left panel shows a scatter plot of test-loss against $\lambda_g$.  The combined v-shaped growth of the minimal achievable test loss with $|\lambda_g|$, along with the existence of points with high test loss at $\lambda_g=0$, together indicate that the condition $\lambda_g=0$ is necessary but not sufficient to achieve low test loss.  Similarly, the top-right panel shows a scatter plot of test-loss against $\lambda_a=0$. The similar v-shaped structure of the scatter indicates that $\lambda_a=0$ alone is necessary but not sufficient to achieve minimal test loss.  

Motivated by these observations, to test whether both conditions together are sufficient to achieve minimal test loss, we attempted to predict test loss via the maximum of a weighted combination $|\lambda_g|$ and $|\lambda_a|$, with learned weights and a bias.  We found an excellent fit (\cref{fig:training_comparison} bottom-left).  The monotonic increase of test-loss with the joint departure of $\lambda_a$ and $\lambda_g$ from $0$, up to a saturation level corresponding to uniform random guessing, indicates that the two conditions $\lambda_a=0$ and $\lambda_g=0$ at initialization are necessary and sufficient for achieving small test loss at the end of training. Moreover, the tightness of the fit in \cref{fig:training_comparison} bottom-left indicates, remarkably, that for a range of low test loss, the final test loss at the {\it end} of training depends on all initial hyperparameters {\it only} through primarily two quantities computed at the {\it beginning} of training: the angle exponent $\lambda_a$ and the gradient exponent $\lambda_g$. Thus, interestingly, combined small departures from the edge of chaos (nonzero $\lambda_a$) and from critical gradient propagation (nonzero $\lambda_g)$ determine final test loss.

\textbf{Conclusion:} In summary, our work provides a novel quantitative geometric theory for forward and backward signal propagation through deep transformers, elucidates $2$ phase transitions and $4$ phases, provides rational guidance for good initialization hyperparameters, and shows remarkably that the final test loss can be predicted by just $2$ exponent functions of these hyperparameters.  Moreover, this entire general theoretical framework could be adapted to analyze other architectures and initialization schemes in future work.  

\section*{Acknowledgements}
S.G. thanks NTT Research, a Schmidt Science Polymath Award, and an NSF CAREER award for funding. A.C., T.N. and X.L.Q. are supported by the National Science
Foundation under grant No. 2111998, and the Simons Foundation.

\bibliography{biblio}
\bibliographystyle{icml2024}

\appendix
\onecolumn

\section{Numerical Computation of the Token Angle Exponent}
\label{app:numerical_token_angle}
We compute the token angle exponent by initializing the tokens with a known initial angle, passing them through a transformer block and then measuring the final token angle and norm.

To initialize the tokens we drawn them from a correlated gaussian distribution so that in expectation they will have norm $X_i \cdot X_i = d = q$ for all $i$ and so that $X_i \cdot X_j = .99d = p$ for all $i \neq j$. We do this so that the tokens are correlated to each other, but the entries of each token are uncorrelated with each other, and indeed $\left(X_i\right)_a$ is uncorrelated with $\left(X_j\right)_b$ (where $a, b$ are indexing the vector of the token) for any $a \neq b$ and for all $i, j$. After passing them through one transformer block we can measure 
\begin{align}
	q' &= \frac{1}{n}\sum_{i=1}^n X_i \cdot X_i \\
	p' &= \frac{2}{n(n-1)} \sum_{i = 1}^n \sum_{j = i+1}^n X_i \cdot X_j.
\end{align}

We may then estimate the token angle expansion exponent as 
\begin{equation}
	\lambda_a = \log \left(\frac{1 - p'/q'}{1-p/q} \right)
\end{equation}

\section{MLP Update Map}
\label{app:mlp_update}
Much of the work of understanding the statistics of pure MLPs has already been accomplished, and so here we will provide information specific to the calculations which underlie the figures and assertions in this paper. In earlier work \citet{poole2016exponential} compute the geometrical behavior of tokens as they pass through an MLP. Given two input vectors (preactivations) $X_1, X_2$ and their norms $|X_1|^2 = |X_2|^2 = q$ and dot product $X_1 \cdot X_2 = p$ \citet{poole2016exponential} compute the norms, $q'$ and dot product $p'$ after one layer of an MLP (again for the preactivations of the next layer). We must do the same, but include a residual connection as well. This is a straightforward extension because for a randomly initialized MLP the preactivation of the next layer is uncorrelated with the preactivation at the previous layer. 

Let $W^1, W^0$ be two weight vectors in the MLP. Then 
Mathematically speaking
\begin{equation}
    \EE\left[(\alpha W^1 \phi( W^0 X_i) + \alphat X_i)\cdot (\alpha W^1 \phi( W^0 X_j) + \alphat X_j)\right] = \alpha^2 \EE\left[ (W^1 \phi( W^0 X_i)) \cdot  W^1 \phi( W^0 X_j) \right] + \alphat^2 X_i \cdot X_j.
\end{equation}

The first term follows from the earlier analysis and there is no cross term because $\EE[W^1] = 0$.

\section{Computing Analytic Lyapunov Exponents}
\subsection{Gradient}
\label{app:gradient_calc}
The analytic calculation of the gradient begins with calculating the second moment of the layer-wise Jacobian matrix for both an attention layer, and an MLP layer, and then combining them to obtain that of a full transformer block. We begin with the computation of an MLP layer. Recall \cref{eq:MLP_action}:
\begin{equation}
    \MLP(X_i) = W^\ell \phi( W^{\ell-1} \cdots W^1\phi(W^0 X_i)).
\end{equation}
We will specialize to the case $\ell = 2$ which can be straightforwardly, if with additional computational effort, extended to the general case. We will also drop the token index $i$ for brevity as the MLP acts independently on each token, and as before we will use greek indices (ex: $\mu$) for the vector components of each token. In this case
\begin{align}
    \frac{\partial}{\partial X_\mu} \MLP(X)_\nu &= \frac{\partial}{\partial X_\mu} \sum_{\alpha}W^2_{\nu\alpha} \phi\left(\sum_{\beta}W^1_{\alpha \beta} \phi\left( \sum_{\gamma} W^0_{\beta\gamma} X_\gamma \right) \right) \\ &= \sum_{\alpha, \beta} W^2_{\nu \alpha} \phi'\left(W^1 \phi\left(W^0X\right)\right)_{\alpha} \cdot W^1_{\alpha\beta} \phi'(W^0X)_\beta \cdot W^0_{\beta\mu} \label{eq:MLP_derivitive_explicit}
\end{align}

At this point we take the outer product and the expectation over the random initialization of $W^k$. Taking the outer product we see that
\begin{equation}
    \EE\left[\left(\frac{\partial}{\partial X_\mu} \MLP(X)_\nu\right) \left(\frac{\partial}{\partial X_\eta} \MLP(X)_\zeta\right)\right] = \frac{\sigma_w^2}{d^2}\delta_{\mu \eta}\delta_{\nu\zeta} \EE \left[ \sum_{\alpha, \beta} \phi'\left(W^1 \phi\left(W^0X\right)\right)_{\alpha}^2 \cdot (W^1_{\alpha\beta})^2 \phi'(W^0X)_\beta^2 \right]
\end{equation}

We take the expectation over $W^1$ now by neglecting the correlation between $(W^1 \phi(W^0X))_\alpha$ and $\left(W^1_{\alpha\beta}\right)^2$ because they only share 1 in $d$ components. Similarly neglect the correlation between $\phi(W^0X)_\gamma$ and $\phi'(W^0 X)_\beta$. Finally we assume that the norm of $\phi(W^0X)$ concentrates which also happens when $d$ is large (see \citet{poole2016exponential} for further discussion). Let it concentrate at $\sqrt{q_1}$ With this in place we note that $(W^0X)_\beta$ and $(W^1 \phi(W^0X))_\alpha$ are gaussian variables with variance $\sigma_w^2|X|^2 / d$ and $\sigma_w^2|\phi(W^0X)|^2 / d$ respectively. Using the fact that $|X| = d$ due to normalization, and that the result no longer depends on the indices $\alpha, \beta$ we have
\begin{equation}
    \EE\left[\left(\frac{\partial}{\partial X_\mu} \MLP(X)_\nu\right) \left(\frac{\partial}{\partial X_\eta} \MLP(X)_\zeta\right)\right] = \frac{\sigma_w^4}{d} \delta_{\mu \eta}\delta_{\nu\zeta} \EE_{z \sim\mathcal{N}(0, \sigma_w^2 q_1)} \left[ \phi'\left(z\right)^2 \right] \EE_{z\sim\mathcal{N}(0, \sigma_w^2)} \left[ \phi'\left(z\right)^2 \right].
\end{equation}
This expression is an easily computeable special function of $\sigma_w$. Notice now that the token-index just goes along for the ride, so the full Jacobian of all the tokens simultaneously has an additional delta function between those indices. Explicitly,
\begin{align}
    \EE\left[\left(\frac{\partial}{\partial X^i_\mu} \MLP(X^k)_\nu\right) \left(\frac{\partial}{\partial X^j_\eta} \MLP(X^l)_\zeta\right)\right] &= \frac{\sigma_w^4}{d} \delta_{\mu \eta}\delta_{\nu\zeta} \delta_{ik}\delta_{jl} \EE_{z \sim\mathcal{N}(0, \sigma_w^2 q_1)} \left[ \phi'\left(z\right)^2 \right] \EE_{z\sim\mathcal{N}(0, \sigma_w^2)} \left[ \phi'\left(z\right)^2 \right].
\end{align}

In the main text we set
\begin{equation}
    f(\sigma_w) \equiv \sigma_w^4 \EE_{z \sim\mathcal{N}(0, \sigma_w^2 q_1)} \left[ \phi'\left(z\right)^2 \right] \EE_{z\sim\mathcal{N}(0, \sigma_w^2)} \left[ \phi'\left(z\right)^2 \right].
\end{equation}
This is justified because $q_1$ is also a function of $\sigma_w^2$ as discussed by \citet{poole2016exponential}. 

Now we turn to the computation in the case of attention. Recalling that attention is 
\begin{gather}
    \Att(X)_i = V \sum_{j}A_{ij} X_j,\\
    A_{ij} = \frac{e^{(Q X_i) \cdot (K X_j)/\sqrt{d}}}{\sum_{k=1}^n e^{(Q X_i) \cdot (K X_k)/\sqrt{d}}}
\end{gather}
Let $Z$ be another set of tokens along with $X$. We first compute
\begin{align}
    \EE\left[\Att(X)^\nu_i \Att(Z)^\zeta_j\right] &= \EE\left[\sum_{\alpha,\beta=1}^d V_{\nu\alpha}V_{\zeta\beta}  \sum_{k,l=1}^n A_{ik}A_{jl} X^\alpha_k Z^\beta_l\right] \\
    &= \frac{1}{d}\delta_{\nu\zeta} \sum_{k,l} \EE\left[A_{ik}A_{jl}\right] X_k \cdot Z_l \\
    &\approx \frac{1}{d}\delta_{\nu\zeta} \sum_{k,l} \frac{e^{\sigma_A^2 (X_i \cdot Z_j)(X_k \cdot Z_l)/d^2}}{\sum_{r, s} e^{\sigma_A^2 (X_i \cdot X_j)(X_r\cdot Z_s)}} (X_k \cdot Z_l)
\end{align}
where we made use of a very similar identity to that derived earlier for the second moment of $A$. To calculate the second moment of the Jacobian we must now take the second derivative of this with respect to $X$ and $Z$ before setting $X = Z$ and then taking $p = q$ to specialize to near the fixed point. Before we begin, note that every time we take a derivative of an exponential we get a term which is $O(1/d)$ because the coefficient of any component of $X$ or $Z$ in the exponential is $O(1/d)$. Therefore to leading order in $d$ we only need to take the derivative of the dot product outside the exponential. This yields

\begin{align}
    \left.\frac{\partial}{\partial X_\mu^i} \frac{\partial}{Z_\eta^j}\EE\left[\Att(X)^\nu_k \Att(Z)^\zeta_l\right] \right|_{X=Z, p=q}&= \left. \frac{1}{d}\delta_{\nu\zeta} \sum_{m,n} \frac{e^{\sigma_A^2 (X_k \cdot Z_l)(X_m \cdot Z_n)/d^2}}{\sum_{r, s} e^{\sigma_A^2 (X_k \cdot X_l)(X_m\cdot Z_n)}} \delta_{im}\delta_{jn}\delta_{\mu\eta} \right|_{X=Z, p=q} \\
    &= \frac{1}{dn^2}\delta_{\nu\zeta} \sum_{m,n} \delta_{im}\delta_{jn}\delta_{\mu\eta} \\
    &= \frac{1}{d n^2} \delta_{\mu\eta} \delta_{\nu \zeta} \mathbf{1}_{i}\mathbf{1}_{j}\mathbf{1}_{k}\mathbf{1}_{l}.
\end{align}
We use the notation that $\mathbf{1}_{i} = 1$, in other words $\mathbf{1}$ is the vector with all ones so that we can explicitly show the token indices. In this analysis we neglected to take into account the normalization of the tokens. The main effect of the normalization is an overall multiplicative factor of $\sqrt{d / |X_i|} = \sqrt{d/q_*}$. Because we consider the second moment we get an overall factor of $d/q_*$, substituting the norm of $X$ at the fixed point. The secondary effect of the normalization is to remove one degree of freedom (along the vector $X_i$) from the gradient. As this is a $O(d^{-1})$ correction we neglect it.

We must now extend both of these results to the cases with the residual connections. Fortunately this is straightforward. Similar to what was discussed in the main text with respect to the update map, the gradients passing through the residual branch are uncorrelated with the gradient passing through the MLP or Attention block. This means that the second moment is just the weighted sum of the second moments of an identity function, and the two formulas we calculated above. Putting these together (abusing notation and allowing $\Att$ and $\MLP$ to represent the functions with residuals, just for this equation) we have that

\begin{equation}
    \EE\left[\left(\frac{\partial \MLP(X)^k_\nu}{\partial X^i_\mu} \right) \left(\frac{\partial \MLP(X^l)_\zeta}{\partial X^j_\eta} \right)\right] = \alphat_M^2 \delta_{ik}\delta_{jl}\delta_{\mu\nu}\delta_{\eta\zeta} + \frac{\alpha_A^2}{q_*} \delta_{\mu \eta}\delta_{\nu\zeta} \delta_{ik}\delta_{jl} f(\sigma_w) = \alphat_M^2 \Id + \frac{\alpha_M^2  f(\sigma_w)}{q_*}\mathcal{A}, \label{eq:final_mlp_app}
\end{equation}
\begin{equation}
    \EE\left[\frac{\partial \Att(X)^\nu_k}{\partial X_\mu^i} \frac{\partial \Att(X)^\zeta_l}{X_\eta^j}\right] = \alphat_A^2 \delta_{ik}\delta_{jl}\delta_{\mu\nu}\delta_{\eta\zeta} + \frac{\alpha_A^2}{q_* n^2}\delta_{\mu\eta}\delta_{\nu\zeta} \mathbf{1}_{i}\mathbf{1}_{j}\mathbf{1}_{k}\mathbf{1}_{l} = \alphat_A^2 \Id + \frac{\alpha_A^2}{q_* n^2} \mathcal{B}. \label{eq:final_attn_app}
\end{equation}

In the final equalities on both lines we suppress the indices and write them in terms of 
\begin{gather}
    \Id_{i\mu j\eta, k\nu l\zeta} = \delta_{ik}\delta_{jl}\delta_{\mu\nu}\delta_{\eta\zeta}, \\
    \mathcal{A}_{i\mu j\eta, k\nu l\zeta} = \delta_{\mu \eta}\delta_{\nu\zeta} \delta_{ik}\delta_{jl}, \\
    \mathcal{B}_{i\mu j\eta, k\nu l\zeta} = \delta_{\mu\eta}\delta_{\nu\zeta} \mathbf{1}_{i}\mathbf{1}_{j}\mathbf{1}_{k}\mathbf{1}_{l}.
\end{gather}

Viewing these as matrices, the comma separates the ``input'' and ``output'' indices. In other words we multiply these symbols by writing them down and summing over middle indices as one does in typical matrix multiplication. 

We now derive the algebra. If the reader is adept at drawing such matrices using tensor network notation then this derivation is a straightforward application of a few minutes and a blackboard.\footnote{See ex: https://tensornetwork.org/diagrams/ for a tutorial on such diagrams} Otherwise the index algebra is considerably more tedious, and we perform this below for the benefit of those not experienced with tensor notation. It is immediately apparent that $\Id$ acts as the identity because it is the outer product of the Jacobin of the identity map with itself. The other terms are as follows: 
\begin{gather}
    (\mathcal{A}\mathcal{A})_{i\mu j\eta, k\nu l\zeta} = \mathcal{A}_{i\mu j\eta, m\beta n\gamma}\mathcal{A}_{m\beta n\gamma, k\nu l\zeta} = \delta_{\mu \eta}\delta_{\beta\gamma}^2 \delta_{im}\delta_{jn}  \delta_{\nu\zeta} \delta_{mk}\delta_{nl} = d \cdot \delta_{\mu \eta} \delta_{ik}\delta_{jl}  \delta_{\nu\zeta} = d\mathcal{A}_{i\mu j \eta, k \nu l \zeta}\\
    (\mathcal{A}\mathcal{B})_{i\mu j\eta, k\nu l\zeta} = \mathcal{A}_{i\mu j\eta, m\beta n\gamma} \mathcal{B}_{m\beta n\gamma, k\nu l\zeta} = \delta_{\mu \eta}\delta_{\beta\gamma}^2 \delta_{im}\delta_{jn} \delta_{\nu\zeta} \mathbf{1}_{m}\mathbf{1}_{n}\mathbf{1}_{k}\mathbf{1}_{l} = d \delta_{\mu \eta}  \delta_{\nu\zeta} \mathbf{1}_{i}\mathbf{1}_{j}\mathbf{1}_{k}\mathbf{1}_{l} = d \mathcal{B}_{i\mu j\eta, k\nu l\zeta} \\
    (\mathcal{B}\mathcal{B})_{i\mu j\eta, k\nu l\zeta} = \mathcal{B}_{i\mu j\eta, m\beta n\gamma} \mathcal{B}_{m\beta n\gamma, k\nu l\zeta} =   \delta_{\mu\eta}\delta^2_{\beta\gamma} \mathbf{1}_{i}\mathbf{1}_{j}\mathbf{1}_{m}\mathbf{1}_{n} \delta_{\nu\zeta} \mathbf{1}_{m}\mathbf{1}_{n}\mathbf{1}_{k}\mathbf{1}_{l} = dn^2 \delta_{\mu\eta}\delta_{\nu\zeta} \mathbf{1}_{i}\mathbf{1}_{j}\mathbf{1}_{k}\mathbf{1}_{l}.
\end{gather}

In these equations we leave the sum over $m, n, \beta, \gamma$ implicit. The remaining product $\mathcal{B}\mathcal{A}$ follows in the same way as for $\mathcal{A}\mathcal{B}$. To obtain the full-block input output Jacobian we now multiply the two terms from \cref{eq:final_mlp_app,eq:final_attn_app} and see that
\begin{align}
    \mathbb{E}\left[\frac{\partial{X}^{t+1}}{\partial X^t}\otimes \frac{\partial{X}^{t+1}}{\partial X^t}\right] &= (\alphat_M^2 \Id + \frac{\alpha_M^2  f(\sigma_w)}{q_*}\mathcal{A})(\alphat_A^2 \Id + \frac{\alpha_A^2}{q_* n^2} \mathcal{B}) \\
    &= \alphat_M^2\alphat_A^2 \Id + \frac{\alpha_M^2 \alphat_A^2 f(\sigma_w)}{q_*} \mathcal{A} + \left(\frac{\alphat_M^2 \alpha_A^2}{q_* n^2} + \frac{\alpha_M^2 \alpha_A^2 f(\sigma_w) d}{q^2_* n^2} \right) \mathcal{B} \\
    &= \alphat_M^2\alphat_A^2 \Id + \frac{\alpha_M^2 \alphat_A^2 f(\sigma_w)}{q_*} \mathcal{A} + \frac{\alpha_A^2}{q_*n^2} \left(\alphat_M^2 + \frac{\alpha_M^2 f(\sigma_w)d}{q_*} \right) \mathcal{B}
\end{align}

This reproduces the result in the paper. Conveniently there is a $3 \times 3$ representation of this Jacobian. The coefficients of $\Id$, $\mathcal{A}$, and $\mathcal{B}$ update in a linear map under multiplication by the second moment of the Jacobian, and therefore the scaling can be easily derived for any finite depth. The top eigenvalue of this matrix would then give the infinite-depth scaling of the size of the Jacobian. The inner product with respect to the vector $B$ can be similarly straightforwardly computed from the explicit form
\begin{equation}
    B_{i \mu j \nu} = \delta_{ij}\delta_{\mu\nu}.
\end{equation}


\section{Setup of Training on Food-101}
\label{app:training_experiment}
We train an vision transformer-like architecture on the Food-101 dataset \cite{bossard14} for our experiments. This dataset comprises of images of different foods in 101 different categories which have size no larger than than $512\times 512$. Our first step is to preprocess these images, resizing them all to $128\times 128$ images, subtracting .5 from the pixel entries, and multiplying them by 2, before converting them into $16 \times 16$ patches. 

The first layer in our image converts these $16^2 \cdot 3$ dimensional patches to 64-dimensional tokens which match our embedding dimension by the means of a linear map. Next we add positional embedding vectors to each patch, which are initialized at random. The zeroth token is special as that token will be the one we look at for our classification label, and has its own positional embedding also randomly initialized. 

We then pass these tokens through the transformer architecture described in the paper with 16-layers of alternating attention and MLP layers. Finally the head of the transformer is a layer-norm operation followed by a linear projection of the zeroth token to obtain the logits for the 101 classes. 

We train with the standard cross-entropy loss with a learning rate of $0.0003$ and a batch size of 256 for a total of 15 epochs, or passes through the training set. We train without dropout regularization using the Adam with all other parameters at their PyTorch defaults. After training we compute the loss on the test set, having preprocessed it identically to the training set. 

\end{document}